\documentclass{article}

\usepackage[numbers]{natbib}
\usepackage[final]{neurips_2019}

\usepackage[utf8]{inputenc} % allow utf-8 input
\usepackage[T1]{fontenc}    % use 8-bit T1 fonts
\usepackage{hyperref}       % hyperlinks
\usepackage{url}            % simple URL typesetting
\usepackage{booktabs}       % professional-quality tables
\usepackage{amsfonts}       % blackboard math symbols
\usepackage{nicefrac}       % compact symbols for 1/2, etc.
\usepackage{microtype}      % microtypography
\usepackage{graphicx}
\title{Separation of target anatomical structure and occlusions in chest radiographs}

\author{%
   Johannes Hofmanninger
   \And
   Sebastian Röhrich
   \And
   Helmut Prosch
   \And
   Georg Langs 
\vspace{4mm}
\\
  Computational Imaging Research Lab\\
  Department of Biomedical Imaging and Image-guided Therapy\\
  Medical University of Vienna, Austria\\
\texttt{www.cir.meduniwien.ac.at}\\
\texttt{johannes.hofmanninger/georg.langs@meduniwien.ac.at} \\
}

\begin{document}

\maketitle

\begin{abstract}
Chest radiographs are commonly performed low-cost exams for screening and diagnosis. However, radiographs are 2D representations of 3D structures causing considerable clutter impeding visual inspection and automated image analysis. Here, we propose a Fully Convolutional Network to suppress, for a specific task, undesired visual structure from radiographs while retaining the relevant image information such as lung-parenchyma. The proposed algorithm creates reconstructed radiographs and ground-truth data from high resolution CT-scans. Results show that removing visual variation that is irrelevant for a classification task improves the performance of a classifier when only limited training data are available. This is particularly relevant because a low number of ground-truth cases is common in medical imaging.
\end{abstract}

\section{Introduction}
Chest radiographs are major radiological examinations for screening and diagnosis of lung pathologies such as atelectasis, effusion, nodules, pneumonia or pneumathorax \cite{Wang2017}. It is a low-cost and fast first-look exam with little radiation exposure compared to Computed Tomography (CT). Thus, radiographic data are abundant and databases of routinely acquired examinations cover a wide range of physiological and pathological variations. At the same time, algorithms improving sensitivity and specificity of findings based on radiographs have major impact on treatment and management of a large number of individual patients. However, in contrast to CT, radiographs are 2D representations of 3D structures causing superimposition of anatomy. The organ of interest is usually overlapped by anatomical structures which are not essential for the examination causing substantial clutter impeding detection \cite{Samei2000}. This is critical in thoracic imaging, as disease or anatomic variation of the superimposed bony thorax or soft tissues and iatrogenic intrathoracic changes can mimic a intrathoracic disease \cite{Gronner1994}.
Highly skilled radiologists learn to separate the relevant variation from spurious noise. However, studies show that lung cancer lesions that are missed in chest radiographs are mostly located behind dense structure such as the ribs \cite{Shah2003}. Both, machine-learning algorithms and inexperienced radiologists may struggle to deal with the variation induced by the multiple superimposed layers (e.g. clavicle, scapula, ribs or heart).

Dual-energy subtraction imaging (DES) is an existing technique to suppress bony structure in radiographs \cite{Vock2009}. Various machine-learning techniques have been proposed to learn the mapping from standard radiographs to DES images \cite{Gusarev2017, Yang2017, Loog2006} to facilitate bone suppression in standard radiographs. However, DES imaging only allows to suppress bony structure leaving shadows induced by liver, heart or breasts unaffected.
Here, we propose a Fully Convolutional Network (FCN) to separate any structure of interest from the overlapping non-essential anatomy utilizing radiographs reconstructed from CT. We utilize high-resolution CT scans to simulate the projected signal of the lung and the non-essential structure independently. We train an FCN to separate these latent signals given the input mixture. Qualitative and quantitative results show, that the FCN is able to reconstruct lung structure in thoracic radiographs while disposing ribs, clavicle, heart, upper mediastinum and organs of the upper abdomen. 

\section{Method}

\begin{figure}[t!]
 \centering
 \includegraphics[width=\textwidth]{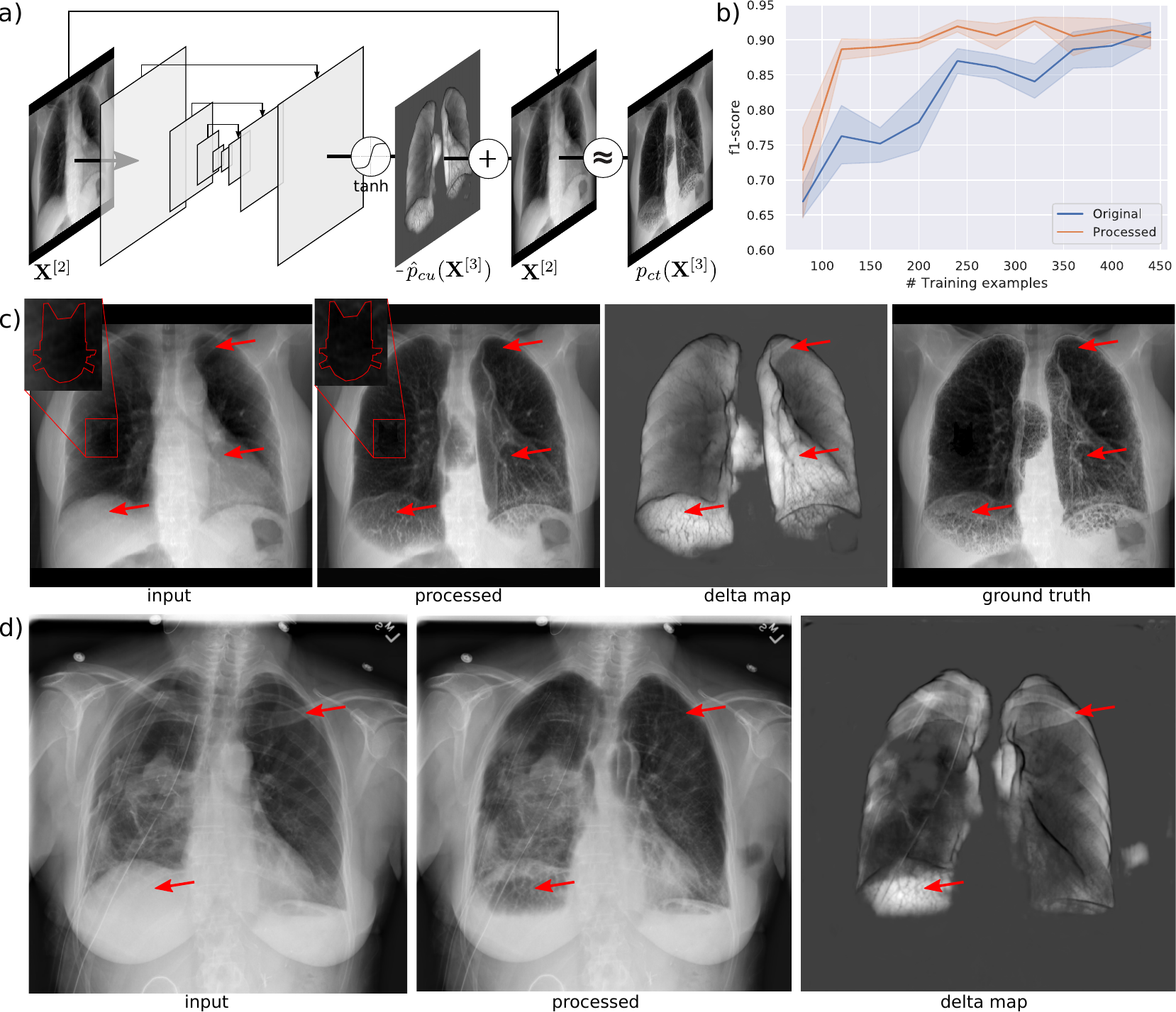}
 \caption{\textbf{Method, quantitative and qualitative results.} \textbf{(a):} Proposed approach to remove undesired image content from a radiograph. A U-Net style FCN learns an additive map to remove the undesired structure from the input image while maintaining the reconstruction of relevant image content \textbf{(b):} Quantitative results showing that processed images (undesired content removed) facilitate the training of a detector when only a limited set of training examples is available. \textbf{(c):} Qualitative result of an image from the hold-out set including the artificial artifact. Red arrows indicate corresponding positions and illustrate the effectiveness of the approach to clear areas behind bone, liver or heart. \textbf{(d):} The network applied on a real radiograph illustrating the transferability of the approach.}
 \label{fig:everything}
\end{figure}

We model a radiograph as a sum of multiple signals or independent components $C$. More formally, we denote $\mathbf{X}^{[2]}$ as a 2D projection of the 3D object $\mathbf{X}^{[3]}$ so that:
\begin{equation}
    \mathbf{X}^{[2]} = \sum_C{p_{c}(\mathbf{X}^{[3]})} \textrm{ \ and \ }
    p : \mathbb{R}^3 \mapsto \mathbb{R}^2
\end{equation}
where $c$ represents an anatomical unit such as an organ or organ part and $p$ is the projection of the unit into the 2-dimensional image space. We further simplify this model to consider only two anatomical units. First, the desired structure $ct$ and second, non-essential structure $cu$ so that an image is 
\begin{equation}
    \mathbf{X}^{[2]} = p_{ct}(\mathbf{X}^{[3]}) + p_{cu}(\mathbf{X}^{[3]})
\end{equation}
and
\begin{equation}
    \mathbf{X}^{[2]} - \hat{p}_{cu}(\mathbf{X}^{[3]}) \approx p_{ct}(\mathbf{X}^{[3]})
\end{equation}
where $\hat{p}_{cu}(\mathbf{X}^{[3]})$ is an estimated delta map holding the undesired variation in the input image learned by an FCN. Here, we use a U-Net-like network \cite{Ronneberger2015} with 5 down- and up-sample steps and learn the parameters by minimizing the Mean-Squared-Error $MSE(\mathbf{X}^{[2]} - p_{cu}(\mathbf{X}^{[3]}),p_{ct}(\mathbf{X}^{[3]}))$. Figure \ref{fig:everything}a outlines the network configuration. During training, the ground-truth image $p_{ct}(\mathbf{X}^{[3]})$ is required. However, creating such images, holding only desired structures from radiographs, is infeasible. We propose to utilize high resolution volumetric CT-scans in combination with a segmentation algorithm of the desired structure (in our case the lung) to generate such images. To this end, we automatically remove structure from the scan that is not present in real radiographs (e.g. the scanner-table) and subsequently generate digitally reconstructed radiographs (DRR) (see Figure \ref{fig:everything}c-\textit{input}) representing $\mathbf{X}^{[2]}$. DRR generation is performed as proposed by Campo et al. \cite{Campo2018} where Hounsfield Units are projected according to the transformation:
\begin{equation}
    \mathbf{X}^{[2]}(x_i,y_i) = \frac{1}{N} \sum^N_{k=1}{e^{\beta\frac{max(\mathbf{X}^{[3]}(x_i,y_j,z_k),-1024)+1024}{1000}}}
\end{equation}
Further, we automatically segment the lung \cite{Hofmanninger2020lungseg}\footnote{https://github.com/JoHof/lungmask} and remove all image information that is in front (ventral) or behind (dorsal) of the lung (e.g. Figure \ref{fig:everything}c-\textit{ground truth}) representing $p_{ct}(\mathbf{X}^{[3]})$. Prior to training, we cropped the projections to the lung area and re-scaled the images to a resolution of 512x512 pixels. In order to facilitate the processing of high-resolution (512x512) images we gradually increased the resolution during training (progressive resizing \cite{Karras2017ProgressiveVariation}) in three steps from 128x128 over 256x256 to 512x512 pixels. 

To quantify the effectiveness of the approach, we artificially induced variation within the lung in the form of a randomly placed low-density area of a particular shape (Figure \ref{fig:everything}-\textit{input} red square). Subsequently, we trained a classifier to distinguish between cases that show this variation and cases which do not. We evaluated, if the classifier would benefit from images that were pre-processed with our approach compared to the original DRR. 

\section{Results}
\textbf{Data: }
We extracted a set of 7400 high-resolution CT scans from 7400 different studies of 4633 patients of the thorax with a maximum voxel-spacing of 1mm in every direction from the Picture Archiving and Communications System (PACS) of a University Hospital. As a first step, we created the DRRs and ground-truth (essential image structure) from all cases. Subsequently, we used 5000 pairs (training-set) of these images to train the network. While the split was performed randomly, we carefully avoided any within-patient overlap between the training- and the hold-out set.

\textbf{Qualitative Results: }
Figure \ref{fig:everything}c illustrates the result of a case that was not used for training. Note, that the network is able to successfully reconstruct fine-grained structure within the lung that was opacified by liver, heart, bone and other tissue. Figure \ref{fig:everything}d shows an example where the network was applied on a real chest radiograph, demonstrating the transferability of the approach. 

\textbf{Quantitative Results: }
For the classification experiment, only cases from the hold-out set were used. We fine-tuned a pre-trained (ImageNet) resnet50 \cite{He2016} to distinguish between cases that show the artificial variation shaped like a cat (50\% of training and testing images showed the cat). We performed training and validation with a varying number of training examples using the remaining cases of the hold-out set for testing. In addition, we repeated each training and testing five times for five epochs each. Training on 120 cases yields an average f1-score of 0.76 (precision=0.67, recall=0.91) on the test set when the unprocessed images are used and 0.89 (precision=0.92, recall=0.86) when the undesired structure is suppressed. Figure \ref{fig:everything}b shows the f1-scores of the classification on the test-set with the number of training images ranging from 80 to 440 using the original synthetic radiographs and the processed images. 

\section{Discussion}
Here, we propose to use deep-learning to suppress undesired visual variation from thoracic radiographs while retaining the structure of interest. Results show that removing visual variation that is irrelevant for a classification task improves the performance of a classifier when a limited training set is available. This is particularly relevant because a low number of ground-truth cases is common in medical imaging. Qualitative results suggest the transferability from synthetic to real radiographs is feasible. The key insight is that models can learn separating compartments in projection imaging modalities from training data, and thereby facilitate the training of detection models for which relatively small amounts of supervised training data is available. A limitation of the approach is that specifics of image acquisition such as patient pose (laying versus standing) can change appearance of certain pathologies such as air-pockets. A further point not addressed in this work is the resolution difference between CT and radiographs, leading to the model not yet fully exploiting the high resolution in radiographs. 

\section{Acknowledgements}
This work was supported by the Austrian Science Fund, FWF I 2714B31, the Austrian National Bank Anniversary Fund OENB 18207, and Siemens Healthineers Digital Health (https://www.siemens-healthineers.com/digital-health-solutions). A GPU used for part of this research was donated by the NVIDIA Corporation.

\bibliography{references}
\medskip

\small

\end{document}